\documentclass[11pt,aps,prl,reprint,superscriptaddress,floatfix]{revtex4-2}
\usepackage[singlespacing]{setspace}
\usepackage{graphicx}
\usepackage{amsmath}
\usepackage{enumitem}
\usepackage{xcolor}
\usepackage{float}
\usepackage{braket}
\usepackage{bbold}
\usepackage{graphicx}
\usepackage{url}
\usepackage[unicode]{hyperref}
\usepackage{cancel}
\usepackage{siunitx}

\begin{document}
	
    \title{Extracting the Anyonic Exchange Phase from Hanbury Brown--Twiss Correlations}
    
	\author{Felix Puster}
	\affiliation{Institut f\"ur Theoretische Physik, Universit\"at Leipzig, Br\"uderstrasse 16, 04103 Leipzig, Germany}

    \author{Matthias Thamm}
    \email{thamm@itp.uni-leipzig.de}
	\affiliation{Institut f\"ur Theoretische Physik, Universit\"at Leipzig, Br\"uderstrasse 16, 04103 Leipzig, Germany}

	\author{Bernd Rosenow}
	\affiliation{Institut f\"ur Theoretische Physik, Universit\"at Leipzig, Br\"uderstrasse 16, 04103 Leipzig, Germany}

    \date{\today}

	\begin{abstract}
	\noindent 
    In recent years, interferometry experiments in fractional quantum Hall devices have reported signatures of a fractional braiding phase for quasiparticles. It was noted, however, that the braiding phase alone does not uniquely determine the exchange phase because of a $\pi$-ambiguity. Here we analyze a Hanbury Brown--Twiss interferometer in a cross geometry that provides direct access to the fractional exchange phase. Using a non-equilibrium Keldysh calculation in an experimentally relevant regime, we show that the exchange phase can be obtained as the phase shift between Aharonov–Bohm oscillations in a single-particle interference current and those in the current cross-correlation arising from two-particle interference.

	\end{abstract}
	\maketitle

The fractional quantum Hall (FQH) effect provides a paradigmatic example of a two-dimensional correlated phase whose quasiparticle excitations are neither fermions nor bosons. These excitations, \emph{anyons}, are expected to exhibit fractional exchange statistics: upon exchanging two identical quasiparticles, the many-body wavefunction acquires a phase factor $\exp(i\theta)$~\cite{Leinaas.1977,Wilczek1982PRL,Laughlin.1983,Halperin.1984,Arovas.1984}. For a recent overview, see Ref.~\cite{Feldman.2021}. While shot-noise measurements established the fractional quasiparticle charge already some time ago~\cite{Kane.1994,Saminadayar.1997,dePicciotto.1997},
obtaining equally direct evidence for fractional exchange statistics has proven more challenging.

     Substantial progress has been made recently. Interferometric experiments have
reported clear signatures of a \emph{braiding} phase $2\theta$
for FQH quasiparticles~\cite{Nakamura.2022,Nakamura.2023,Lee.2023,Glidic.2023,Ruelle.2023,Samuelson.28.03.2024,Kim.27.12.2024,Werkmeister.2025,Ruelle.2025},
building on theoretical proposals for Fabry--P\'erot interferometers
and related geometries~\cite{C.Chamon.1997,Halperin.2011,Rosenow.2020,Han.2016}
as well as on ``anyon collider'' approaches based on partitioning and noise measurements~\cite{Rosenow.2016,Bartolomei.2020,Schiller.2023,Jonckheere.2023,Thamm.2024,Iyer.2024}. In these devices, quasiparticles propagate along chiral edge channels and tunnel between edges at quantum point contacts (QPCs), enabling interference phenomena. However, it was pointed out recently that the braiding phase accessible in such
measurements determines the exchange phase only modulo $\pi$~\cite{Read.2024}. In particular, a determination of $\theta$ up to $\pi$ cannot distinguish $\theta=0$ from $\theta=\pi$.

       In this Letter we analyze an interferometer that extracts the exchange phase $\theta$ without $
       \pi$-ambiguity from Hanbury Brown--Twiss (HBT) correlations~\cite{Henny1999Science,Oliver1999Science}. The exchange phase has appeared previously in HBT-type proposals for anyonic edge states~\cite{Safi.2001,Kim.2005,Kim.2006,Samuelsson.2004,Campagnano.2012,Campagnano.2013}, but typically entangled with additional non-universal ingredients, making it difficult to isolate $\theta$ experimentally. Ref.~\cite{Vishveshwara.2003} considered a closely related geometry, but focused
on time-resolved correlations and did not emphasize that in quantum Hall interferometers the statistical phase appears additively with an unavoidable Aharonov--Bohm (AB) phase. Recently, Ref.~\cite{Kivelson.18.03.2024} proposed a modified Fabry--P\'erot interferometer in which an antidot embedded in a QPC
provides an alternative route to accessing $e^{i\theta}$ by controlling the anyon occupancy of the antidot. Our approach is complementary: it relies on two-particle interference and yields direct access to the statistical phase through an HBT contribution to current cross-correlations~\cite{Buttiker1992PRB}, with an experimental
complexity comparable to that of existing two-particle interferometers.

\begin{figure}
        \centering
        \includegraphics[width=\linewidth]{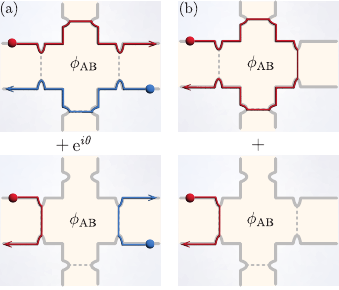}
        \caption{\label{fig:interference}
        Interfering processes contributing to
two-particle interference in the current cross-correlation (a) and to
single-particle interference in the current (b). Both acquire an
Aharonov--Bohm phase from the enclosed flux.
 In (a), the two outgoing states
differ by exchanging the two quasiparticles, yielding an additional statistical
phase $\theta$ and therefore shifting the AB oscillations relative to (b).       
    }
    \end{figure}

The central idea is to use the AB phase as a reference. We compare (i) an interference contribution to a current, governed by single-quasiparticle paths, to (ii) an interference contribution to a zero-frequency current cross-correlation, governed by two-quasiparticle processes of the Hanbury Brown--Twiss type~\cite{Brown.1954,BROWN.1956}.  In the latter case, the
two interfering processes lead to outgoing states that differ by an exchange of the two quasiparticles, so that the AB oscillations of the cross-correlation are shifted by $\theta$ relative to the AB oscillations of the current, see   Fig.~\ref{fig:interference}. Because $\theta$ is obtained as a \emph{relative} phase shift between two
oscillatory signals measured in the same device, slow drifts of the absolute AB
phase (e.g., from area breathing) largely cancel. Measuring
this phase shift therefore yields $\theta$ directly.

    \begin{figure}
        \centering
        \includegraphics[width=\linewidth]{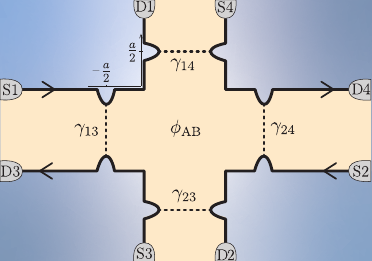}
        \caption{\label{fig:setup}
        Schematic of the cross-geometry interferometer. Four chiral edges are connected pairwise to their neighbors by QPCs, enabling
both single-particle and two-particle interference. On each edge the two QPCs
are separated by a distance $a$, and the tunneling amplitudes are denoted by
$\gamma_{nm}$. The enclosed magnetic flux produces the Aharonov--Bohm phase
$\phi_{AB}$.}
    \end{figure}
    
    In the following, we focus on Laughlin states~\cite{Laughlin.1983}, i.e., filling fractions $\nu=1/m$ with odd, positive integer $m$, for which there is a single
edge mode carrying a single type of quasiparticle with exchange phase
$\theta=\pi\nu$.

    \paragraph{Setup.} 
    The setup consists of four chiral edge channels arranged in a cross geometry
(cf.~Fig.~\ref{fig:setup}). Such an open, gate-defined geometry is compatible with modern high-mobility
platforms in which bulk--edge Coulomb coupling can be strongly reduced, e.g.,
GaAs/AlGaAs heterostructures incorporating additional screening wells~\cite{Nakamura.2022}
and graphene devices with proximal graphite gates/back gates~\cite{Samuelson.28.03.2024,Werkmeister.2025}.
This helps avoid interaction-driven phase shifts familiar from Fabry--P\'erot
interferometers~\cite{Halperin.2011}.   Each edge is coupled to each of its two neighboring
edges by a QPC. On a given edge, the two tunneling points are separated by a
distance $a$. We parametrize each edge by a coordinate $x$ that runs from
$x=-\infty$ at the source contact to $x=+\infty$ at the drain contact, with
$x=0$ chosen at the inner corner. With this convention, the two tunneling points
on each edge lie at $x=\mp a/2$, as indicated in Fig.~\ref{fig:setup} for edge~1.
    
   The edge dynamics are described by a chiral Luttinger liquid appropriate to the
fractional quantum Hall effect~\cite{Wen.1990,Wen.1991}. In the absence of
tunneling, the four edges are governed by
    \begin{equation}
        H_0 = \frac{\hbar v}{4\pi \nu} \sum_{i=1}^4 \int dx [\partial_x \phi_i(x)]^2,
    \end{equation}
    where $v$ is the edge-mode velocity and the boson fields $\phi_i$ encode the
anyon statistics through the equal-time commutator
$[\phi_i(x),\phi_j(y)]= i\theta\,\delta_{ij}\,\text{sign}(x-y)$.
To describe tunneling, we introduce quasiparticle operators
$\psi_i(x,t)=(2\pi)^{-\nu/2}\kappa_i\,e^{i\phi_i(x,t)}$ that annihilate an anyon
on edge~$i$. The Klein factors $\kappa_i$ ensure the correct mutual statistics
between operators acting on different edges; equivalently, one may imagine that
the edges are connected at infinity to form a common coordinate system.
Their algebra can be parameterized by an antisymmetric matrix $\alpha$~\cite{Guyon.2002,Altland.2015},
    \begin{equation}
        \kappa_i \kappa_j = \kappa_j \kappa_i\, e^{i\pi\nu\alpha_{ij}},\quad \overline{\kappa}_i \kappa_j= \kappa_j  \overline{\kappa}_i \, e^{-i\pi\nu\alpha_{ij}},
    \end{equation}
and $\kappa_i \overline{\kappa}_i=1$. 
Although the final results are independent of the choice of global ordering, we
let the common coordinate system start at source~S1, which yields
    \begin{align}
        \alpha &=  \left( \begin{matrix}
            0 & 1 & 1& 1 \\
            -1 & 0 & 1 & -1 \\
            -1 & -1 & 0 & -1 \\ 
            -1 & 1  & 1 & 0
        \end{matrix} \right) \ .  \label{eq:AlphaKlein}
    \end{align}  
    We now define tunneling operators that transfer one anyon from edge $n$ to edge $m$,
    \begin{equation}
        O^+_{nm}(t)=\gamma_{nm}\psi^\dagger_m(x_{mn},t)\psi^{\phantom{\dagger}}_n(x_{nm},t)\ ,
    \end{equation}
 and $O^-_{nm}(t)=[O^+_{nm}(t)]^\dagger$, where $\gamma_{nm}$ is the tunneling
amplitude and $x_{ij}$ denotes the point on edge $i$ connected to edge $j$.
In   the interaction picture the tunneling Hamiltonian reads
    \begin{equation}
        H_{\text{tun}}(t)=\sum_{n=1}^2\sum_{m=3}^4 O_{nm}^+(t)e^{-i\frac{e^*}{\hbar}(V_n-V_m) t}+\text{h.c.}\,,
    \end{equation}
with fractional charge $e^*=\nu e$.

In the following, we compute the AB and statistical contributions to the
interference phase in the current cross-correlation, as well as the AB-phase
dependence of the single-particle interference contribution to the current,
perturbatively to leading order in the tunneling amplitudes $\gamma$.
A caveat is that quasiparticle tunneling at each QPC is an RG-relevant
perturbation~\cite{Kane.1992}.    At zero temperature, the leading-order
tunneling current involves an energy integral over the product of tunneling-in
and tunneling-out densities of states~\cite{Kane.1992}. With a finite voltage
drop at a QPC, the corresponding singularities are separated; if the voltage
drop vanishes, they coincide and the perturbative current diverges.
For consecutive tunneling through two QPCs, this implies that a voltage drop at each
QPC is required for zero-temperature perturbation theory to remain finite~\cite{Kane.2003b}.    
Here, the interference terms involve products of densities of states from all four edges, and at $T=0$, they can diverge even when each QPC has a voltage drop, if opposite edges are biased identically so that singularities align. 
 We will therefore introduce a small voltage difference $\Delta V$ between the edges to
regularize the zero-temperature expressions.

    \paragraph{Current Cross-Correlations.}
We consider a bias configuration in which quasiparticles tunnel at the QPCs such
that current is collected at drains $D3$ and $D4$. The corresponding current
operators can be written as
    \begin{equation}\label{eq:currents}
        I_3=\frac{\nu e^2}{h}V_3+I_{13}+I_{23},\ I_4=\frac{\nu e^2}{h}V_4+I_{14}+I_{24},
    \end{equation}
    where $I_{nm}$ denotes the tunneling contribution from edge $n$ to edge $m$,
    \begin{align}
        I_{nm}(t)&=\frac{ie^*}{\hbar} \left[O^+_{nm}(t)e^{-i\frac{e^*}{\hbar}(V_n-V_m) t}- \text{h.c.}\right].
    \end{align}
Following the original HBT idea, we expect that the current cross-correlations
between drains $D3$ and $D4$ contain a two-particle interference contribution
in which the two interfering processes differ by a single exchange of
quasiparticles and hence acquire an additional statistical phase
(cf.~Fig.~\ref{fig:interference}(a)).
We therefore compute the symmetrized zero-frequency cross-correlation,
to leading order in the tunneling amplitudes,
    \begin{align}
        S&=\frac{1}{2}\int_{-\infty}^{\infty}dt\ \langle \Delta I_3(t) \Delta I_4(0)\rangle+\langle \Delta I_4(0)\Delta I_3(t)\rangle, 
    \end{align}
with $\Delta I_\alpha=I_\alpha-\langle I_\alpha \rangle$. We evaluate the non-equilibrium expectation values using the Keldysh formalism and, from now on, set $V_3=0$ as the reference potential.
 Then we obtain
    \begin{align}\label{eq:8_pointsnoise_on_contour_expand_simplify}
        &S(V,T)\approx\sum_{n,m=1}^2\int_{-\infty}^{\infty}dt \,\Bigg\{-\langle I_{n3}\rangle\langle I_{m4}\rangle\nonumber\\
        &+\frac{(e^*)^2}{\hbar^4}\text{Re}\Bigg[\sum_{s_1,s_2=\pm}\sum_{n',m'=1}^2 s_1 s_2 \sum_{\sigma_1,\sigma_2=\pm}\sigma_1\sigma_2\nonumber\\
        &\times \int_{-\infty}^{\infty}dt^{\prime}\int_{-\infty}^{\infty}dt^{\prime\prime} e^{s_1 \frac{i}{\hbar}(e^*V_{n'} t^{\prime}-e^*V_n t)}e^{s_2 \frac{i}{\hbar}(e^*V_{m'}-e^*V_{4}) t^{\prime\prime}}\nonumber\\
        &\times \left\langle T_C O^{s_1}_{n3}(t_{-})O^{s_2}_{m4}(0_{+}) O_{n'3}^{-s_1}(t^{\prime}_{\sigma_1}) O_{m'4}^{-s_2}(t^{\prime\prime}_{\sigma_2})\right\rangle\Bigg]\Bigg\} \; .
    \end{align}
Here $s_i=\pm$ label the tunneling direction and $\sigma_i=\pm$ are Keldysh
branch indices. In the second term, one can distinguish four classes of
contributions: (i) $m\neq n$ with $n'=n$ and $m'=m$, which factorize into
$\langle I_{n3}\rangle\langle I_{m4}\rangle$ and are therefore canceled by the
explicit subtraction in Eq.~\eqref{eq:8_pointsnoise_on_contour_expand_simplify};
(ii) $m=n$ with $n'=n$ and $m'=n$, which would factorize for non-interacting
electrons, but in the FQH case, contain interaction effects calculated in
Refs.~\cite{Safi.2001,Kim.2005,Kim.2006} and have no AB-phase dependence;    
(iii) $m\neq n$ with $n'=m$ and $m'=n$, which yield the AB-dependent part of
$\langle I_{n3}I_{m4}\rangle$; and    
(iv) $m=n$ with $n'\neq n$ and $m'=n'$, which yield the AB-dependent part of
$\langle I_{n3}I_{n4}\rangle$.
    
It is convenient to express the result in terms of the contour-ordered Green
function
    \begin{align}\label{eq:8_points_G_contour_anyon}
        G&(x-x',t_{\sigma_1},t_{\sigma_2}')=e^{i\pi\nu\text{sgn}(x-x')[\frac{1-\sigma_{12}}{2}]}\nonumber\\
        &\times\left[\frac{\frac{k_B T}{\hbar v}\epsilon}{2\text{sin}(\pi \frac{k_B T}{\hbar v}[\epsilon-i\sigma_{12}(x-x')+i\sigma_{12}v(t-t')])}\right]^\nu,
    \end{align}
where $\sigma_{12}=\frac{1}{2}(\sigma_2-\sigma_1+\text{sgn}(t-t')[\sigma_1+\sigma_2])$
and $\epsilon$ is a UV cutoff. The prefactor in Eq.~(10)
collects the Klein-factor phases generated when applying Wick's theorem on the
Keldysh contour; we absorb it into $G$ for notational convenience.
 We emphasize
that this phase does not originate from exchanging the two operators in the
definition of $G$.
   
Unlike pulsed/time-domain collider settings where finite wavepacket shape can
control short-time correlations~\cite{Thamm.2024,Iyer.2024}, we focus on
steady-state zero-frequency correlations under dc bias, for which such
microscopic shaping affects visibilities but not the extracted phase shift. To extract the exchange phase, we choose the voltage configuration
$V_2=V$, $V_1=V+\Delta V$, and $V_4=-\Delta V$, with $V>\Delta V>0$.
The large bias $V$ realizes the two-particle interference picture, while the
small offset $\Delta V$ regularizes the zero-temperature expressions, as
discussed above. Focusing on the AB-dependent part $S_{AB}(\phi_{AB})$ (terms of
type (iii) and (iv)), we obtain
    \begin{align}\label{eq:S_AB}
        &S_{AB}(V,\Delta V,T)=(e^*)^2\frac{e^*V}{\hbar}\frac{\left(\frac{\epsilon}{\hbar v}\right)^{4\nu}|\gamma_{13}^*\gamma_{14}\gamma_{24}^*\gamma_{23}|}{(e^* V)^{4(1-\nu)}}\nonumber\\
        &\times \cos(\phi_{AB}+\pi\nu)\tilde{S}(V,\Delta V,T)+\mathcal{O}\left(a\frac{e^*V}{\hbar v}\right),
    \end{align}
where $\tilde{S}$ is a dimensionless integral. At $T=0$ and for large bias,
$\tilde{S}$ depends only on $\Delta V/V$ and, to leading order, is
    \begin{align}\label{eq:8_points_noise_together_T0_final_result_regime}
       \tilde{S}(V,\Delta V,0)&=4\frac{(2\pi)^{3-4\nu}}{\Gamma(\nu)^3}\frac{\Gamma(1-2\nu)}{\Gamma(1-\nu)}\left(\frac{\Delta V}{V}\right)^{-1+2\nu}.
    \end{align}
     
 In agreement with the picture discussed in the introduction, the AB-phase
dependence of the HBT correlations is a cosine in which the AB phase is shifted
additively by the statistical contribution.   For the special voltage choice
above, there are no further corrections to this phase shift arising from finite
temperature or from the finite separation of the QPCs, as demonstrated
numerically in the supplemental material~\cite{Supplement}.

    \paragraph{Current.}
To extract the exchange phase from the HBT correlations, we need an independent
determination of the AB phase. We obtain it from the single-particle
interference contribution to the current at drain $D3$ when a large bias is
applied to source $S1$ (cf.~Fig.~\ref{fig:interference}(b)).
Concretely, we consider the voltage configuration $V_1=V$ and
$V_2=V_4=-\Delta V$, with $V>\Delta V>0$, at zero temperature, and decompose the
current as $I_3=I_{3,0}+I_{3,AB}(\phi_{AB})$ (see Eq.~\eqref{eq:currents}).
We find   
    \begin{align}\label{eq:I_AB}
        I_{3,AB}&(V,\Delta V,T=0)=e^*\frac{e^* V}{\hbar}\frac{\left(\frac{\epsilon}{\hbar v}\right)^{4\nu}|\gamma_{13}^*\gamma_{14}\gamma_{24}^*\gamma_{23}|}{(e^* V)^{4(1-\nu)}}\nonumber\\
        &\times \cos(\phi_{AB})\tilde{I}\left(\zeta \right)+\mathcal{O}\left(a\frac{e^*V}{\hbar v}\right),
    \end{align}
where $\tilde{I}$ is a dimensionless integral. For $\Delta V /V \ll 1$ its leading behavior is
    \begin{align}\label{eq:8_points_interference_terms_grt_anyon_T0_Green}
        \tilde{I}\left(\frac{\Delta V}{V}\right)&=2\frac{(2\pi)^{3-4\nu}}{\Gamma(\nu)^3}\frac{\Gamma(2-3\nu)}{\Gamma(2-2\nu)}\left(\frac{\Delta V}{V}\right)^{-2+3\nu}.
    \end{align}
    Thus, in this regime, the interference contribution to the current is governed
by $\cos(\phi_{AB})$ (up to the small corrections indicated explicitly).
 A general expression at finite temperature is given in the supplemental
material~\cite{Supplement}. Finite $a$ and finite $T$ can additionally induce a
small phase shift of the current oscillations, discussed below.

\begin{figure}
        \centering
        \includegraphics[width=1\linewidth]{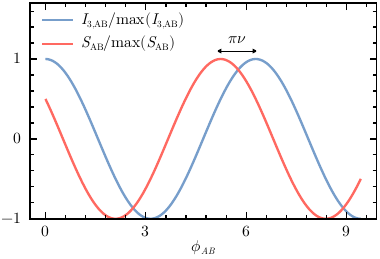}
        \caption{\label{fig:shift}
        Normalized Aharonov--Bohm oscillations of the
interference current and of the AB-dependent current cross-correlation in their
respective voltage configurations. The relative phase shift equals the exchange
phase $\theta=\pi\nu$, shown here for $\nu=1/3$. }
    \end{figure}

\begin{figure}
        \centering
        \includegraphics[width=1\linewidth]{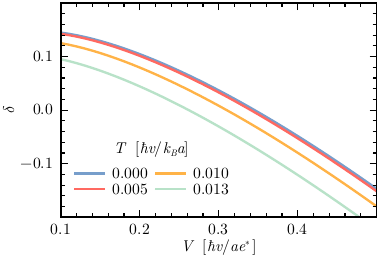}
        \caption{\label{fig:current_shift}
        Additional phase shift $\delta$ of the
single-particle interference current as a function of bias $V$ for several
temperatures $T$. For $\nu=1/3$ and fixed dimensionless
$e^*a\Delta V/(\hbar v)=0.06$, we extract $\delta$ from
$I_{3,AB}\propto\cos(\phi_{AB}+\delta)$ in the regime of
Eq.~\eqref{eq:I_AB}. }
    \end{figure}

    \paragraph{Discussion.}
Plotting the normalized AB oscillations of the current and of the current
cross-correlations in their respective voltage configurations reveals the
anyonic exchange phase as a phase shift between the two oscillations
(cf.~Fig.~\ref{fig:shift}). Crucially, extracting $\theta$ from the relative shift between $I_{3,AB}$ and
$S_{AB}$ provides intrinsic common-mode rejection: global phase drifts that
shift both oscillations similarly do not affect the inferred statistical phase.
In the ideal limit, $I_{3,AB}\propto \cos(\phi_{AB})$
whereas $S_{AB}\propto \cos(\phi_{AB}+\theta)$; in practice, the interference
current may acquire a small additional phase shift $\delta$ due to finite
temperature and the finite separation $a$ of the QPCs. This correction remains
much smaller than the exchange phase under experimentally relevant conditions,
as illustrated by our numerical evaluation of $I_{3,AB}$ and the extracted
$\delta$ in Fig.~\ref{fig:current_shift}.
   
For $\nu=1/3$, $T=10\,\mathrm{mK}$, $V=\SI{60}{\micro\volt}$, edge velocity
$v=10^5\,\mathrm{m/s}$, and separation $a=\SI{1}{\micro\meter}$, we have
$ae^*V/\hbar v\approx 0.30$ and $k_BTa/\hbar v\approx 0.013$.
There are two competing corrections associated with the finite separation $a$, and an additional small correction at finite temperature,
which can be partially balanced by optimizing the small bias $\Delta V$.
Numerically, we find that for the parameters stated above, the corrections are small for
$\Delta V=V/5$. In this case, the residual shift is approximately
$\delta\approx -0.010\,\pi$, much smaller than the exchange phase
$\theta=\pi/3$. Thus, our setup enables a direct and unambiguous determination
of the fractional exchange phase.
    
\emph{Edge structure.} Edge reconstruction (or coupling to additional edge modes)
can renormalize the exponent in quasiparticle correlation functions, effectively
replacing $\nu$ by a nonuniversal exponent~\cite{Rosenow.2002}. This
changes the voltage and temperature dependence of visibilities but does not affect
the phase shift between single- and two-particle interference, which
remains set by $\theta$.

\emph{Coulomb effects.} Coulomb coupling can modify interference phases and may
lead to Coulomb-dominated behavior in Fabry--P\'erot devices~\cite{Halperin.2011},
but the present cross geometry is comparatively open and should be well screened
by the four edge channels. Screening is further enhanced in modern GaAs
heterostructures with dedicated screening wells~\cite{Nakamura.2022} and in
graphene interferometers with nearby graphite gates/back gates~\cite{Samuelson.28.03.2024}.
      

    \paragraph{Generalized Fano factor.} 
    As an additional characterization, one may consider the ratio of the amplitudes
of the AB oscillations of the current cross-correlation and of the current in
the regime $\Delta V/V\ll 1$. Up to an overall factor of the fractional charge,
all nonuniversal and dimensional quantities cancel. Dividing this ratio by the
elementary charge defines a \emph{generalized Fano factor},
$F=\nu\,\tilde{S}/\tilde{I}$.
At zero temperature, and up to corrections of order $(\Delta V/V)^{2-3\nu}$ and
higher, we find
\begin{align}\label{eq:Fano}
        F&=\frac{\nu}{2}\frac{(1-2\nu)\Gamma(1-2\nu)^2}{\Gamma(1-\nu)\Gamma(2-3\nu)}\left(\frac{\Delta V}{V}\right)^{1-\nu}.
    \end{align}
Thus, the filling fraction $\nu$ controls both the prefactor and the exponent
of the voltage dependence. Since all quantities entering
Eq.~\eqref{eq:Fano} are experimentally accessible, this relation provides a
useful consistency check for the applicability of the perturbative theory.

    \paragraph{Conclusion.} 
We have analyzed a cross-geometry interferometer for Laughlin quasiparticles and
computed, to leading order in the tunneling amplitudes, the AB-dependent parts
of (i) a zero-frequency current cross-correlation and (ii) a current.
For the cross-correlation, the AB oscillations acquire an additive shift by one
fractional exchange phase, consistent with an HBT-type two-particle
interference process. The single-particle interference contribution to the
current, by contrast, depends (up to small corrections) on the bare AB phase.
As a result, the exchange phase becomes directly accessible experimentally as
the phase shift between the AB oscillations of the cross-correlation and of the
current. We have shown numerically that corrections to the current phase from
finite temperature and finite QPC separation remain much smaller than the
exchange phase for realistic parameters. This enables an unambiguous
determination of the exchange statistics of Laughlin quasiparticles.

\ \\
    \begin{acknowledgments}
    F.P.\@ acknowledges support by the Deutsche Forschungsgemeinschaft (DFG) under the Grant No.\@ 406116891 within the Research Training Group RTG 2522/1.
    \end{acknowledgments}

    \clearpage

    \section{Supplement}  
    \subsection{General solution for the current cross-correlations}
\begin{figure}
        \centering
        \includegraphics[width=\linewidth]{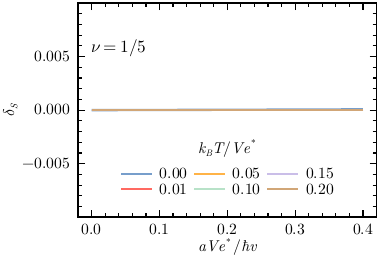}
        \caption{
In general, the AB-dependent part of the current
cross-correlation can be written as
$S_{AB}\propto\cos(\phi_{AB}+\pi\nu+\delta_S)$. The numerical results show that
for the voltage configuration used in the main text the additional phase shift
$\delta_S$ vanishes even at finite temperature and for finite separation of the
tunneling points. Parameters: $\tilde{V}=10$ and $\Delta\tilde{V}=2.5$. }
        \label{fig:noise_shift}
    \end{figure}

In the main text, we focused on a particular bias configuration in which the
AB-dependent part of the current cross-correlations takes a simple form.
Here, we provide additional details on how the Keldysh expression, Eq.~\eqref{eq:8_pointsnoise_on_contour_expand_simplify}, is evaluated.
Both terms of Eq.~\eqref{eq:8_pointsnoise_on_contour_expand_simplify} can be expressed in terms of the contour-ordered Green function
\begin{align}\label{eq:8_points_G_contour_anyon_supplemental}
        G(x&-x',t_{\sigma_1},t_{\sigma_2}')=e^{i\pi\nu\text{sgn}(x-x')[\frac{1-\sigma_{12}}{2}]}\nonumber\\
        &\times\left[\frac{\frac{k_B T}{\hbar v}\epsilon}{2\text{sin}(\pi \frac{k_B T}{\hbar v}[\epsilon-i\sigma_{12}(x-x')+i\sigma_{12}v(t-t')])}\right]^\nu,
    \end{align}
    where $\sigma_{12}=\frac{1}{2}(\sigma_2-\sigma_1+\text{sgn}(t-t')[\sigma_1+\sigma_2])$. We then find 
    \begin{align}\label{eq:8_points__nmmn_general_simplify_s1}
        &S_3=2\frac{(e^*)^2}{\hbar^4}|\gamma_{13}^*\gamma_{14}\gamma_{24}^*\gamma_{23}|\nonumber\\
        &\times \text{Re}\Bigg[\sum_{\sigma_1,\sigma_2=\pm}\sum_{\substack{n,m=1\\ n\neq m}}^2\int_{-\infty}^{\infty}dt\ \sigma_1\sigma_2\int_{-\infty}^{\infty}dt^{\prime}\int_{-\infty}^{\infty}dt^{\prime\prime}  \nonumber\\
        &\times\cos[(\phi_{AB}+\pi\nu)(n-m)-\tilde{V}_n (t-t^{\prime\prime})+\tilde{V}_{m} t^{\prime}-\tilde{V}_4 t^{\prime\prime}]\nonumber\\  
        &\times G(-a(n-m),t^{\prime}_{\sigma_1},0_{+}) G(a(n-m),t^{\prime}_{\sigma_1},t_{-}) \nonumber\\
        & G(-a(n-m),t^{\prime\prime}_{\sigma_2},t_{-})G(a(n-m),t^{\prime\prime}_{\sigma_2},0_{+})\vphantom{\sum_{n,m\in\{1,2\}}\int_a^b} \Bigg] \ , \vspace{-0.2cm}
    \end{align}
    and  
    \begin{align}\label{eq:8_points_noise_on_contour_nnll_sum_s1}
        &S_4=-2\frac{(e^*)^2}{\hbar^4}|\gamma_{13}^*\gamma_{14}\gamma_{24}^*\gamma_{23}|\nonumber\\
        &\times\text{Re}\Bigg[ \sum_{\sigma_1,\sigma_2=\pm}\sum_{\substack{n,l=1\\ n\neq l}}^2\int_{-\infty}^{\infty}dt\ \sigma_1\sigma_2\int_{-\infty}^{\infty}dt^{\prime}\int_{-\infty}^{\infty}dt^{\prime\prime}  \nonumber\\
        &\times\text{cos}[(n-l)(\phi_{AB}+\pi\nu)-\tilde{V}_n t+\tilde{V}_l(t^{\prime}-t^{\prime\prime})+\tilde{V}_4 t^{\prime\prime}]\nonumber\\
        &\times G(a(n-l),t_{-},0_{+})G(-a(n-l),t_{-},t^{\prime}_{\sigma_1}) \nonumber\\
        &\times  G(-a(n-l),t^{\prime}_{\sigma_1},t^{\prime\prime}_{\sigma_2})G(a(n-l),0_{+},t^{\prime\prime}_{\sigma_2})\vphantom{\sum_{n,m\in\{1,2\}}\int_a^b} \Bigg].
    \end{align}
To evaluate these integrals efficiently at finite QPC separation $a$, it is
useful (for $\epsilon\ll 1$) to employ
        \begin{align}
        G(x,t_+,t'_+)&\approx\begin{cases}
            G(x,t_-,t'_+) &,x>0\\
            G(x,t_+,t'_-) &,x<0
        \end{cases} \\ \intertext{}
        G(x,t_-,t'_-)&\approx\begin{cases}
            G(x,t_+,t'_-) &,x>0\\
            G(x,t_-,t'_+) &,x<0
        \end{cases}\ ,
    \end{align}
which amounts to replacing $\sigma_{12}$ in Eq.~\ref{eq:8_points_G_contour_anyon_supplemental}
by $\tilde{\sigma}_{12}=\frac{1}{2}(\sigma_2-\sigma_1+\text{sgn}(x)[\sigma_1+\sigma_2])$.
A convenient next step is to insert the Fourier representation
    \begin{align}
        &G(x,t_{\sigma_1},t_{\sigma_2}')=\nonumber\\
        &e^{i\pi\nu\text{sgn}(x)[\frac{1-\tilde{\sigma}_{12}}{2}]}\left[\frac{\frac{k_B T \epsilon}{v\hbar}}{2\text{sin}(\pi \frac{k_B T}{\hbar}[\frac{\epsilon}{v}-i\tilde{\sigma}_{12}\frac{x}{v}+i\tilde{\sigma}_{12}(t-t')])}\right]^\nu\nonumber\\
        &=\frac{2e^{i\pi\nu\text{sgn}(x)[\frac{1-\tilde{\sigma}_{12}}{2}]}}{\epsilon^{-\nu}v^\nu}\int_{-\infty}^{\infty}\frac{d\omega}{2\pi} e^{-\omega[\frac{\tilde{\sigma}_{12}\epsilon-i x+iv(t-t')}{v}]} \Omega_r(\tilde{\sigma}_{12}\omega)\label{eq:8_points_G_contour_anyon_Fourier_space},
    \end{align}
    with 
    \begin{equation}
        \Omega_r(\omega)=\begin{cases}
        \Theta(\omega)\frac{\pi}{(2\pi)^\nu\Gamma(\nu)}|\omega|^{-1+\nu},& T=0\\
        \frac{\pi \left(\frac{k_B T}{\hbar}\right)^{\nu -1} \frac{e^{\frac{\hbar \omega}{2 k_B T}}}{\cosh \left(\frac{\hbar \omega}{k_B T}\right)-\cos (\pi  \nu )}}{2 \Gamma (\nu ) \Gamma \left(1-\frac{\nu }{2}-\frac{i \hbar \omega}{2 \pi k_B T}\right) \Gamma \left(1-\frac{\nu }{2}+\frac{i \hbar \omega}{2 \pi k_B T}\right)} & T>0
        \end{cases}\ ,
    \end{equation}
where the subscript $r$ indicates that $\Omega_r$ is real-valued. While the
finite-temperature expression is convenient for numerical evaluation, at $T=0$
one obtains a particularly simple form for $S_{AB}$, neglecting corrections of
order $a\tilde{V}/v$:
\begin{align}\label{eq:8_points_noise_together_T0_final_result}
        &S_{AB}(V_1,V_2,V_4,T=0)=32\frac{(e^*)^2}{\hbar^4}\text{sign}(V_1)\text{sign}(V_2-V_4)\nonumber\\
        &\times \left[\frac{\pi}{(2\pi)^\nu\Gamma(\nu)}\right]^4\left(\frac{\epsilon}{v}\right)^{4\nu}|\gamma_{13}^*\gamma_{14}\gamma_{24}^*\gamma_{23}|\cos(\phi_{AB}+\pi\nu)\nonumber\\
        &\times\int_{f_{\text{low}}(\tilde{V}_1,\tilde{V}_2,\tilde{V}_3)}^{f_{\text{up}}(\tilde{V}_1,\tilde{V}_2,\tilde{V}_3)}\frac{d\omega}{2\pi}(|\omega||\omega-\tilde{V}_1||\omega-\tilde{V}_2||\omega-\tilde{V}_4|)^{-1+\nu}\nonumber\\
        &+\mathcal{O}\left(\gamma^6,a\frac{e^*V}{\hbar v}\right),
    \end{align}
    with $f_{\text{low}}(\tilde{V}_1,\tilde{V}_2,\tilde{V}_3)=\text{max}[\text{min}(\tilde{V}_1,0),\text{min}(\tilde{V}_2,\tilde{V}_4)]$, and $f_{\text{up}}(\tilde{V}_1,\tilde{V}_2,\tilde{V}_3)=\text{min}[\text{max}(\tilde{V}_1,0),\text{max}(\tilde{V}_2,\tilde{V}_4)]$. 
We can also check numerically that there are no corrections to the phase shift
at finite temperature and finite separation between the tunneling points if one
chooses the voltage configuration from the main text, $V_2=V$, $V_1=V+\Delta V$,
$V_4=-\Delta V$, with $V\gg\Delta V>0$.
Figure~\ref{fig:noise_shift} shows the
additional phase shift $\delta_S$ defined by
$S_{AB}\propto\cos(\phi_{AB}+\pi\nu+\delta_S)$ for $\tilde{V}=10$,
$\Delta\tilde{V}=2.5$, and $\nu=1/5$.

    \vspace*{.3cm}
    \subsection{General solution for the AB-oscillations of the current}
The AB-dependent part of the current can be written in terms of the same
contour-ordered Green functions as
\begin{align}\label{eq:8_points_current_grt}
        &I_{3,AB}=\nonumber\\
        &-2\frac{e^*}{\hbar^4}\text{Re}\Bigg[\sum_{\substack{n,m=1\\ n\neq m}}^2 e^{i\pi\nu(n-m)}e^{i\phi_{AB}(n-m)}|\gamma_{13}^*\gamma_{14}\gamma_{24}^*\gamma_{23}| \nonumber\\
        &\times \sum_{\sigma_1,\sigma_2,\sigma_3=\pm}\sigma_1\sigma_2\sigma_3\int_{-\infty}^{\infty}dt \int_{-\infty}^{\infty}dt^{\prime} \int_{-\infty}^{\infty}dt^{\prime\prime}\nonumber\\
        &\times e^{i\tilde{V}_nt-i\tilde{V}_m(t^{\prime\prime}-t^{\prime})-i\tilde{V}_4(t-t^{\prime\prime})} \nonumber\\
        &\times G(a(n-m),0_{+},t_{\sigma_1}) G(-a(n-m),0_{-},t^{\prime}_{\sigma_2})\nonumber\\
        &\times G(a(n-m),t_{\sigma_1},t^{\prime\prime}_{\sigma_3}) G(-a(n-m),t^{\prime}_{\sigma_2},t^{\prime\prime}_{\sigma_3}) \Bigg].
    \end{align}
    At zero temperature, this reduces to
\begin{align}\label{eq:8_points_interference_terms_grt_anyon_T0_Green_final}
       &I_{3,AB}= 2\frac{e^*}{\hbar^4}|\gamma_{13}^*\gamma_{14}\gamma_{24}^*\gamma_{23}|\left[\frac{2\pi}{(2\pi)^\nu\Gamma(\nu)}\right]^4\left(\frac{\epsilon}{v}\right)^{4\nu}\nonumber\\
       &\times \text{Re}\Bigg\{e^{i\phi_{AB}}\!\int_{-\infty}^{\infty}\frac{d\omega}{2\pi} \left[|\omega|\,|\omega-\tilde{V}_1|\,|\omega-\tilde{V}_2|\,|\omega-\tilde{V}_4|\right]^{-1+\nu} \nonumber\\
       &\times e^{i \frac{a}{v} \left(-V_1- V_2-V_4\right)}e^{i 4\frac{a}{v}  \omega_1}\nonumber\\
   &\times \left[\Theta\left(\omega
   -\tilde{V}_2\right)-e^{i\pi\nu}\Theta\left(\tilde{V}_2-\omega\right) \right]\nonumber\\
        &\times
   \left[\Theta\left(\omega-\tilde{V}_4\right)-e^{i\pi\nu}
   \Theta\left(\tilde{V}_4-\omega\right)\right]\nonumber\\
        &\times \Big[\Theta\left(\tilde{V}_1-\omega\right)\Theta\left(\omega \right)-\Theta\left(-\omega\right)\Theta\left(\omega-\tilde{V}_1\right)\Big]\Bigg\}\nonumber\\
        &+\mathcal{O}\left(\gamma^6\right).
    \end{align}

\end{document}